\documentclass[pdflatex,sn-nature, oneside]{sn-jnl}
\usepackage{graphicx}%
\usepackage{multirow}%
\usepackage{amsmath,amssymb,amsfonts}%
\usepackage{amsthm}%
\usepackage{caption}
\usepackage{manyfoot}%
\usepackage{braket}
\usepackage{mathrsfs}%
\usepackage[title]{appendix}%
\usepackage{xcolor}%
\usepackage{textcomp}%
\usepackage{manyfoot}%
\usepackage{booktabs}%
\usepackage{algorithm}%
\usepackage{algorithmicx}%
\usepackage{algpseudocode}%
\usepackage{listings}%

\renewcommand{\thefigure}{\arabic{figure}}
\DeclareCaptionFont{snfig}{\fontsize{8}{9.5}\selectfont}

\begin{document}

\title{Attosecond quantum spectroscopy with entangled photon pairs}

\author[1]{Zijian Lyu}
\author[1]{Fengxiao Sun}
\author[2]{Sili Yi}
\author[1]{Jingze Li}
\author[1]{Haodong Liu}
\author[1]{Qiongyi He}
\author[1]{Qihuang Gong}
\author[2,*]{Misha Ivanov}
\author[1,3,4,*]{Yunquan Liu}

\affil[1]{State Key Laboratory for Mesoscopic Physics  and Frontiers Science Center for Nano-optoelectronics, School of Physics, Peking University, Beijing, 100871, China}
\affil[2]{Max Born Institute, Max-Born Stra$\beta$e 2A, D-12489 Berlin, Germany}
\affil[3]{Collaborative Innovation Center of Extreme Optics, Shanxi University, Taiyuan, 030006, China}
\affil[4]{Peking University Yangtze Delta Institute of Optoelectronics, Nantong, Jiangsu, 226010, China}

\affil[*]{Mikhail.Ivanov@mbi-berlin.de}
\affil[*]{Yunquan.Liu@pku.edu.cn}


\abstract
{Bright squeezed light from parametric down-conversion in the infrared (IR) frequency range has triggered the emergence of attosecond quantum optics -- a new research field at the interface of quantum optics, strong-field physics, and attosecond technology. Two challenges arise at this interface: transferring  quantum features of the IR light sources to the ultraviolet (UV) and extreme ultraviolet (XUV) frequency range via strong-field nonlinearities, and  exploiting quantum optical properties of the nonlinear optical response as a new probe in ultrafast dynamics. Here, we address both by driving high-harmonic generation (HHG) in solids with entangled photon pairs either in degenerate or non-degenerate frequency modes. In the degenerate mode, single-shot measurements of harmonics up to the 10th order reveal strong photon bunching whose $g^{(2)}$ first grows and then decreases with the harmonic order. We show that this behavior tracks different microscopic mechanisms responsible for harmonic emission, demonstrating the potential of attosecond quantum optical spectroscopy. In the non-degenerate case, the harmonics retain quantum-induced correlations, verified by wavelength-resolved second-order cross-correlation maps. 
Our findings demonstrate transfer of quantum photon correlations into the XUV domain and open a pathway toward quantum-enhanced attosecond spectroscopy and control of ultrafast dynamics in solids. 
}

\maketitle

\newpage

\section*{Main}

In quantum electrodynamics, radiation by a classical current generates Glauber coherent states \cite{glauber1963coherent} of the electromagnetic field \cite{scully1997quantum}. These states most closely resemble classical electromagnetic fields, exhibiting Poissonian photon statistics and minimal quantum noise \cite{glauber1963coherent}. In contrast, quantum states of light 
can display such nonclassical features as photon anti-bunching, squeezing or entanglement \cite{grynberg2010introduction}. 
These coherence and correlation properties of  the radiation are encoded in the various order correlation functions 
$g^{(n)}$ (e.g. $g^{(2)}$)
in both temporal and spectral domains and are related to the physical mechanisms responsible for light generation \cite{loudon2000quantum}.
\\
\indent 
In attosecond technology, high-harmonic generation (HHG) has played a central role as a source of highly coherent and widely tunable, attosecond-scale radiation. It arises from the highly nonlinear interaction of intense femtosecond laser fields with matter, be it atoms, molecules, liquids, or solids. 
Almost universally since its discovery \cite{ferray1988multiple,mcpherson1987studies}, high-harmonic generation has been described within the semiclassical framework \cite{lewenstein1994theory, corkum1993plasma,schafer1993above}, in which the driving light field is treated classically and the emitted harmonics are expected to follow suit, i.e. emerging in the Glauber coherent states within the quantum optical framework. This expectation was  natural given the enormous number of photons (e.g. $10^{13}$) incident in the coherent state describing the driving laser field, and given the typically very large number (e.g. $10^6$) of the generated harmonic photons. 
\\ \indent
However, recent studies of quantum properties of harmonic generation driven by classical  light \cite{tsatrafyllis2017high, tsatrafyllis2019quantum,lewenstein2021generation, theidel2024evidence,theidel2025observation} have demonstrated that this expectation is not always met, opening a new frontier of strong-field 
quantum optics. Post-processing involving conditioning the driving field measured after the generating medium on the harmonic emission could transform the transmitted field into a Schr$\mathrm{\ddot{o}}$dinger cat-like state \cite{lewenstein2021generation, lamprou2025nonlinear}. Perhaps even more strikingly, conventional HHG in semiconductors was sometimes found to exhibit nonclassical features \cite{theidel2024evidence,theidel2025observation}. Theoretically, scenarios involving ground-state depletion \cite{stammer2024entanglement}, medium pre-excitation \cite{rivera2024squeezed}, or the back action of the generated harmonics on the generating medium  \cite{yi2025generation, gonoskov2024nonclassical} have been shown to
yield non-classical states of generated light. 
Importantly, these non-classical properties may also stem from correlated many-body dynamics \cite{pizzi2023light,lange2024electron} and excitonic effects \cite{lange2025excitonic}, pointing to
the sensitivity of the quantum properties of the generated light to the detailed physical mechanisms responsible for its generation.
\\
\indent Another approach to generating nonclassical radiation is to drive the medium with intense quantum light \cite{even2023photon,gorlach2023high}, taking advantage of the technological breakthroughs in generating very bright quantum states of light known as bright squeezed vacuum (BSV) \cite{sh2012superbunched,iskhakov2012polarization, iskhakov2016heralded}. Bright squeezed vacuum (BSV) has recently been employed to drive solids-state HHG, reaching nonperturbative  harmonic yields comparable to those achieved with classical strong fields \cite{rasputnyi2024high}. Using high-order frequency mixing with the combination of a strong classical field and a weak BSV light offers another promising route towards generating new frequency lines that could potentially inherit quantum properties from the BSV \cite{lemieux2025photon, tzur2025measuring}. 

However, the quest for imprinting quantum properties on high-harmonic light has so far left the spectroscopic potential of quantum high-harmonic generation largely unexplored. In particular, in the BSV-driven harmonic generation in solids, what can one learn from the behavior of the time domain second-order correlation function $g^{(2)}(\tau=0)$ across different harmonic orders? Can one relate it to electron dynamics inside the crystal, and to the microscopic mechanisms of harmonic emission driven by quantum light?   

The use of BSV light sources offers an important opportunity to address these questions, as this light can be used in both degenerate and non-degenerate modes. In the degenerate regime, where the two photons are generated at the same frequency, the resulting BSV light exhibits squeezing without entanglement. In contrast, when two non-degenerate Schmidt modes are generated (two-mode BSV), they form spectrally separated bright entangled photon pairs (BEP) possessing both squeezing and genuine bipartite entanglement. The fundamentally important question whether the harmonics generated from such two-mode squeezed state preserve entangled features at higher photon energies has remained unexplored so far. We  address these important questions in this work.

Turning to the quantum-optical high harmonic spectroscopy using the degenerate mode, we follow the photon bunching as a function of 
the harmonic order. We find that the $g^{(2)}$ correlation function first grows and then decreases with the harmonic order. We show that this behavior tracks different microscopic mechanisms responsible for harmonic emission, demonstrating the potential of attosecond quantum optical spectroscopy. In the non-degenerate case,
where the solid is strongly driven by entangled photon pairs, we find that the harmonics retain quantum-induced correlations, verified by wavelength-resolved second-order cross-correlation maps. 

 \begin{figure}[p]
\centering
\includegraphics[width=1\textwidth]{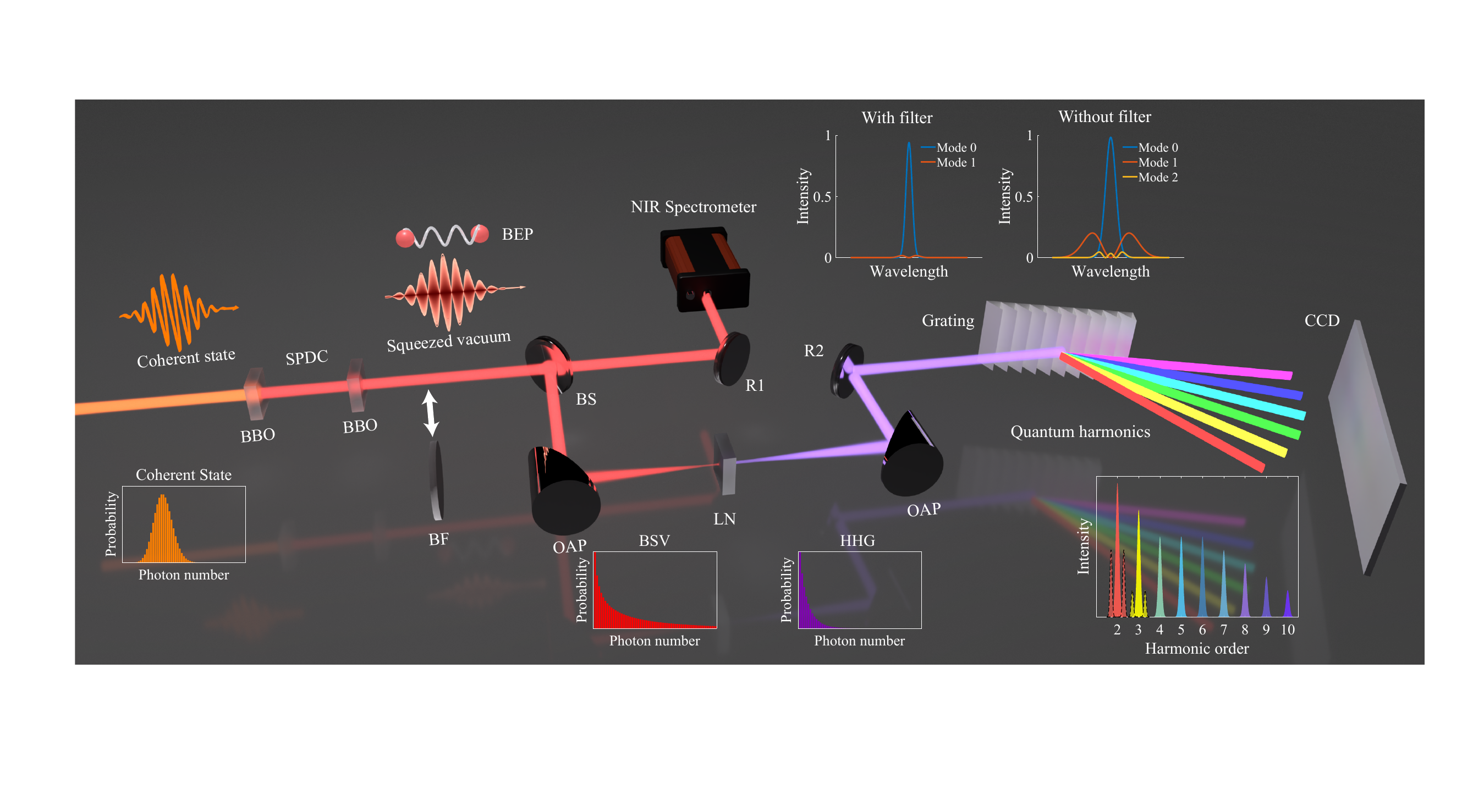}
\caption*{\small \textbf{Figure 1. Schematic of the experimental setup for quantum-light-driven high-harmonic generation in solids.} A coherent pump in a double-BBO geometry undergoes high gain SPDC to produce BSV, which contains a degenerate Schmidt mode (single-mode squeezed state) and non-degenerate modes forming BEP (two-mode squeezed state). A band-pass filter (BF) can isolate the degenerate mode; without filtering, the output remains spectrally multi-mode. The quantum light is then focused by off-axis parabolic mirrors (OAP) into a LN crystal to drive HHG. A beam splitter (BS) can direct a portion of the beam to near-infrared (NIR) spectrometer for characterizing the incident quantum field. The generated harmonics are spectrally resolved by a diffraction grating and recorded shot-by-shot with a CCD. The harmonic spectra reveal the modal composition of the driving field: single-mode BSV leads to narrow, single-peaked harmonics, while multi-mode BSV with BEP sidebands introduces additional peaks in the lower harmonic orders. Quantum features of the driving field are imprinted onto the emitted harmonics. Insets (bottom) show photon number statistics of the coherent state, BSV and HHG.  (R1,R2: reflect mirror)   } \label{fig1}
\end{figure}
\begin{figure}[p]
\raggedright
\includegraphics[width=1\textwidth]{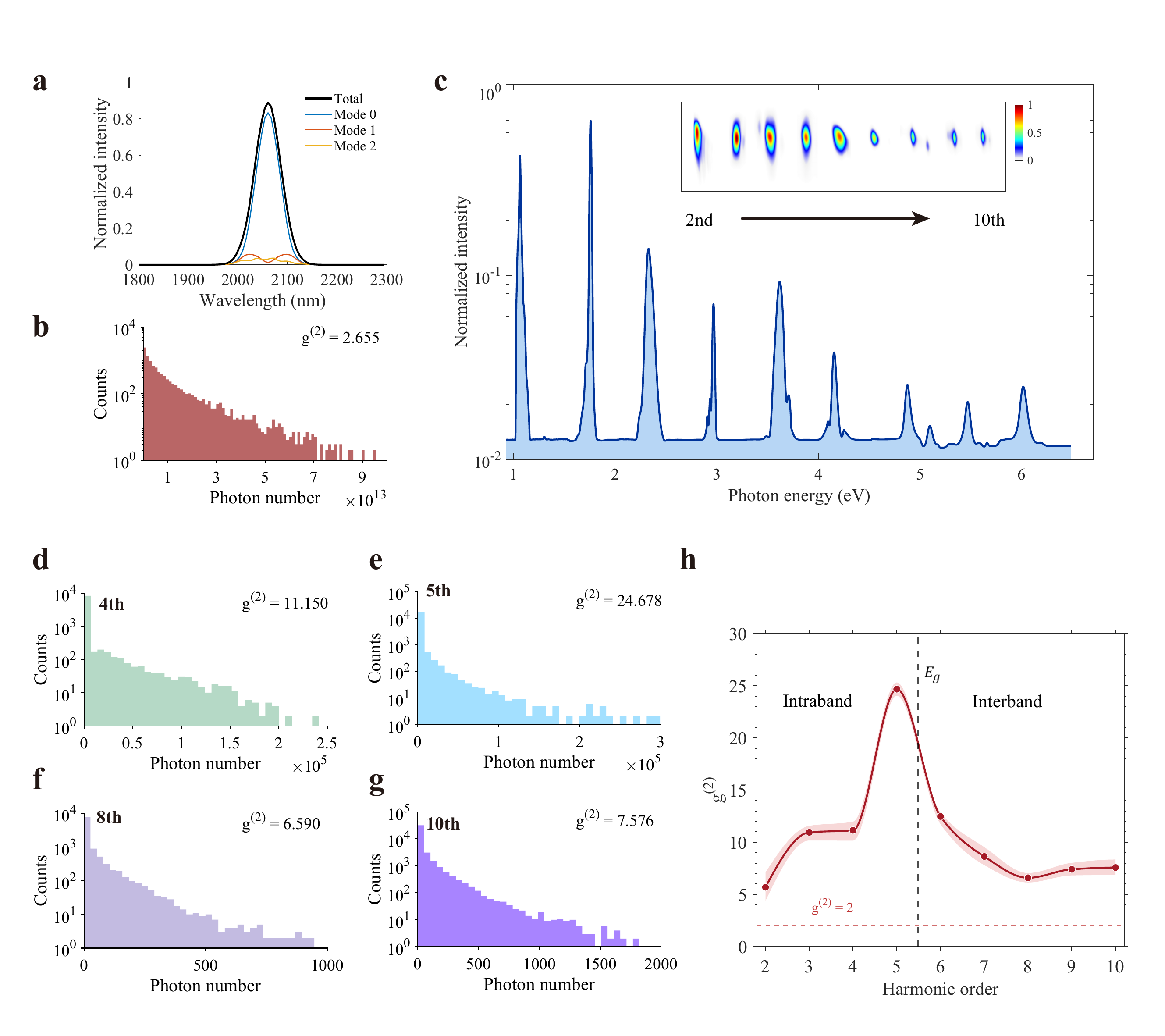}
\caption*{\small  \textbf{Figure 2. Single-mode BSV-driven HHG.} \textbf{a,} Spectrum of the  spectrally filtered BSV and its Schmidt mode decomposition. The total spectrum (black) is dominated by the first (degenerate) mode (blue), with negligible contribution from high-order modes (orange, yellow), confirming that spectral filtering yields a near-single-mode BSV centered at $2060$ nm. \textbf{b,} Photon number distribution of the filtered BSV. The time domain second-order correlation $g^{(2)}(\tau=0)=\langle N^2\rangle/\langle N\rangle^2$, is $2.655$, close to the theoretical value 3 for single-mode BSV. \textbf{c.}  Average harmonic spectra for the 2nd-10th harmonics and the corresponding 2D spectra. (Inset) \textbf{d, e, f, g,} Photon number distribution for the 4th, 5th, 8th and 10th harmonics, with $g^{(2)}=11.150, 24.678, 6.590,$ and $7.576$, respectively, evidencing strong photon bunching. \textbf{h,} $g^{(2)}$ versus harmonic order reveals a non-monotonic trend: increasing from the 2rd to the 5th harmonic, peaking at the 5th, then decreasing at higher orders. The peak aligns closely with the bandgap energy $E_g$ of lithium niobate, which marks the transition from harmonics dominated by intraband currents to those governed by interband polarization. This behavior reflects the interplay between quantum photon statistics and microscopic emission mechanisms.    
}\label{fig2}
\end{figure}

\begin{figure}[p]
\includegraphics[width=1.2\textwidth,height=1.2\textheight,keepaspectratio]{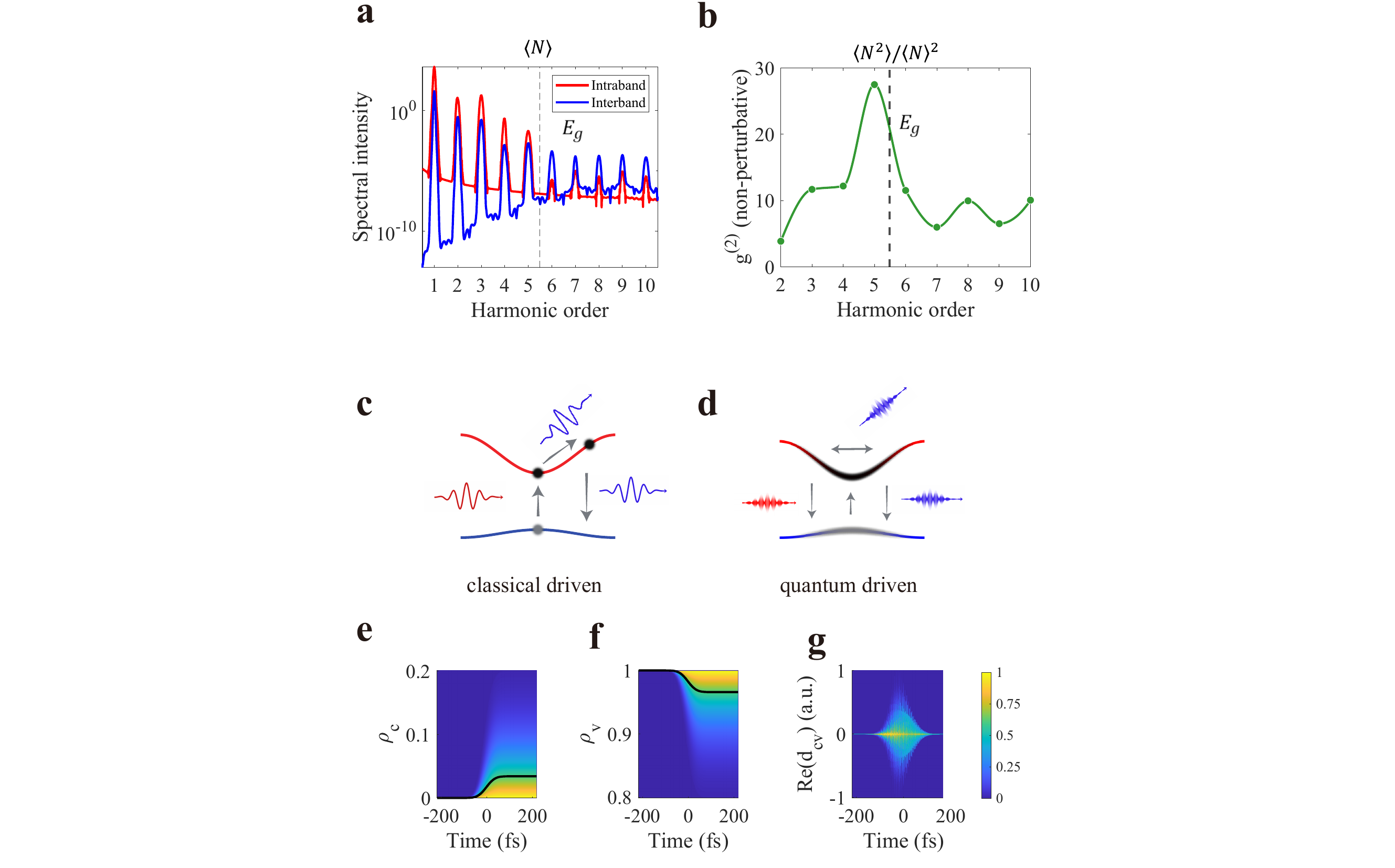}
\caption*{\small \textbf{Figure 3. Quantum-driven HHG dynamics.} \textbf{a,} Spectral contributions $\langle N\rangle$ from intraband (red) and interband (blue) processes to the total HHG yield, obtained via SBE simulation. Harmonics below the bandgap are dominated by the intraband current, whereas those above the bandgap are governed by the interband polarization. \textbf{b,} Numerical simulations based on SBE reproduce the experimental measurement of $g^{(2)}=\langle N^2\rangle/\langle N\rangle ^2$, showing that the microscopic electron  dynamics are sensitive to the quantum fluctuations of the driving field.\textbf{c,d} Conceptual illustration of HHG driven by classical (coherent) and quantum (BSV) light. In the quantum-driven scenario, large quantum fluctuations lead to a significantly broadened carrier distribution in the band structure. \textbf{e-g,} Time evolution of \textbf{e,} conduction band electron population $\rho_e$, \textbf{f,} valence band population $\rho_v$ and \textbf{g,} interband polarization $\Re(d_{cv})$. In \textbf{e} and \textbf{f}, the black curves indicate the results under coherent driving at the same average  intensity for comparison with the quantum-driven case. } 
\label{fig3}
\end{figure}

We generate bright squeezed vacuum (BSV) state light via strongly pumped spontaneous parametric down conversion, SPDC. 
Figure 1 schematically illustrates the generation and control of bright squeezed vacuum (BSV) in high-harmonic generation (HHG). When a band-pass filter is applied, the degenerate mode (Mode 0) is isolated, yielding a single-mode squeezed state. In the absence of spectral filtering, the output remains multi-mode, keeping higher-order Schmidt modes. 

The modal composition of the quantum light strongly influences both the spectral shape and the statistical features of the resulting harmonics.  When driven by single-mode BSV, the resulting harmonic spectrum features a single narrow peak for each order. In contrast, in the multi-mode case, when the harmonic emission is driven by entangled photon pairs, we obtain more complex harmonic spectra. In particular, the sidebands emerge at low orders, reflecting the spectral structure of the entangled modes.  

Experimentally, we begin by investigating HHG driven by single-mode BSV. By applying spatial and spectral filtering, we isolate the first (degenerate) Schmidt mode, yielding a single-mode BSV centered at $2060 \ \mathrm{nm} $ with an intensity approaching $1.2\ \mathrm{TW/cm^2} $.
Schmidt-mode analysis  of the driving field spectrum (see Methods) confirms its near single-mode nature (Fig. 2a): Mode 0 dominates, while higher-order modes (i.e., Mode 1 and 2) are much weaker.  
The measured photon number statistics yield the second-order correlation of $g^{(2)}= 2.655$, close to the theoretical value of $3$. 
Using this single-mode quantum light as the driving field,
we successfully observe  high-harmonic generation with harmonics up to the $10\mathrm{th}$ order in an $x$-cut lithium niobate (LN) crystal, entering the nonperturbative regime (Fig 2c). \\

The averaged spectral intensity reflects the mean photon number $\langle N\rangle $ of each harmonic, serving as a first-order characterization of the emission. To access photon correlations, we perform single-shot measurements of the photon number statistics of individual harmonic orders, with the typical results for the $\mathrm{4th}, \ \mathrm{5th},\ \mathrm{8th}$ and $\mathrm{10th}$ harmonics shown in Fig. 2d, e, f, and g. As expected \cite{rasputnyi2024high}, these harmonics exhibit clear deviations from Poissonian statistics. The observed $g^{(2)}$ values are significantly higher than that of the driving field, indicating stronger photon bunching in the emitted harmonics. Their dependence on harmonic order is plotted in Fig. 2h. Rather than increasing monotonically as expected in the perturbative regime \cite{spasibko2017multiphoton}, we observe a non-monotonic trend: $g^{(2)}$ increases from the $\mathrm{2nd}$ to the $\mathrm{5th}$ order, then gradually decreases for higher orders. Notably, this turning point in the photon statistics coincides with the transition of the harmonic
photon energy across the material bandgap. 

To understand the origin of this $g^{(2)}$ behavior, we perform numerical simulations based on the semiconductor Bloch equations (SBE),
using single-mode BSV as the driving field (see Methods). As shown in Fig. 3a, the calculated spectral intensity $\langle N\rangle$ reveals a transition between the two dominant HHG mechanisms near the bandgap, inter-band and intra-band. 
However, separating the two intensity contributions in the 
combined intensity signal in an experiment is 
very challenging. Our results suggest that measuring photon correlations can help. Indeed, the simulated $g^{(2)}$ reproduces the experimental trend well (Fig.3b) and shows that the harmonics generated by the intraband current and interband polarization exhibit markedly different $g^{(2)}$ behavior. Thus, higher-order correlations such as $g^{(2)}$ provide both additional information and a valuable criterion for identifying the physical process underlying the emission.  
\indent 
Next, we investigate the underlying dynamics  to gain insight into the microscopic origin of the observed high $g^{(2)}$ values and photon superbunching. An essential property of squeezed light is that the fluctuations in the one quadrature are reduced (squeezing) while being enhanced in the conjugate quadrature (anti-squeezing) to satisfy the Heisenberg uncertainty relation. In strong-field processes such as solid-state HHG, the influence of the increased 
field fluctuations in the anti-squeezed quadrature becomes particularly significant. As illustrated in Fig. 3c,d, compared  with classical coherent driving, the larger field fluctuations of the BSV modifies the electrons dynamics and produces a noticeably broader distribution of the carrier distribution in $k$-space. This is reflected in the temporal evolution of the band populations $\rho_c(t)$ and $\rho_v(t)$ (Fig. 3e,f), as well as in the interband polarization $d_{cv}(t)$ (Fig. 3g), all of which exhibit a substantially broadened distribution under BSV excitation. Importantly, these broadened distributions induce fluctuations in both intraband and interband emission processes, leading to harmonics with strongly bunched photon statistics. In this way, the pronounced shot-to-shot quantum noise of the driving BSV is mapped onto fluctuations of the microscopic current and polarization, and is ultimately encoded in the photon number statistics of the generated harmonics. 

This microscopic picture also helps explain the observed non-monotonic order dependence of $g^{(2)}$ associated with the crossover between intraband- and interband-dominated emission. For each crystal momentum $k$, we write the two-band state as $\ket{\psi_k}=a_v(k,t)\ket{v}+a_c(k,t)\ket{c}$, where $a_v$ and $a_c$ are the valence- and conduction-band electron amplitudes. Near the ground state ($a_{\rm{c}0}=0, \ a_{\rm{v}0}\simeq1$), we denote the $k$-resolved excitation amplitude by a small parameter $\epsilon=a_{\rm{c}}(k,t)$. The conduction- and valence-band populations then scale as $\rho_{cc}(k,t)=|a_{\rm{c}}|^2\approx\epsilon^2$, $\rho_{vv}(k,t)=|a_{\rm{v}}|^2\approx 1-\epsilon^2$, while the interband coherence scales as $\rho_{cv}(k,t)=a_{\rm{c}} a_{\rm{v}}^*\simeq \epsilon\sqrt{1-\epsilon^2}\approx \epsilon+\rm{O(\epsilon^3)}$. Consequently, the microscopic source terms entering the two emission channels have different sensitivities to the fluctuations of the excitation amplitude: the intraband current is predominantly population-driven and therefore scales with $\rho_{\rm{cc}}(\rho_{\rm{vv}})\sim\epsilon^2$, whereas the interband polarization is coherence-driven and scales with $\rho_{\rm{cv}}\sim \epsilon$. 

As a result, shot-to-shot variations of the driving field are more strongly converted into fluctuations of the intraband contribution, giving rise to larger bunching at lower orders. 
At higher orders, as the emission shifts toward the above-gap, where interband channel becomes increasingly important, the harmonic photon statistics become less sensitive to the fluctuations of the driving field, leading to a decrease of $g^{(2)}$ with harmonic order. This intraband-to-interband crossover therefore naturally produces the observed rise, peak, and subsequently decline of $g^{(2)}$ across the harmonic spectrum.      


\indent We now move beyond the single-mode case and consider the BSV field with the first few spectral Schmidt modes, i.e., a two-mode BSV that manifest as bright entangled photon pairs (BEP). This configuration can be conveniently realized by simply removing the spectral filter from the optical path. 
For the two-mode squeezed state, a distinctive feature is the entanglement between the optical modes. Here we ask how these nonclassical inter-mode correlations influence the harmonic spectroscopy, in particular the shot-to-shot fluctuations and spectral correlation structure.  

The measured spectrum of the multi-mode driving light can be decomposed into orthogonal Schmidt modes. Fig. 4b shows the spectra of the three dominant modes (Mode 0-2). Under our experimental conditions, the Schmidt eigenvalues are strongly concentrated in these three modes, so the driving field effectively occupies a low-dimensional Schmidt subspace that is relevant for realizing a BEP state. Higher-order modes beyond Mode 2 contribute only a minor fraction of the total energy, therefore the nonlinear HHG response is expected to be governed predominately by the first few modes. The measured photon number distribution of this multi-mode field (Fig. 4c) exhibits a reduced $g^{(2)}=1.669$, compared with the single-mode case. To quantify how different spectral components fluctuate jointly from shot to shot, we evaluate the wavelength-resolved second-order cross-correlation function \cite{glauber1963quantum}, 
\begin{equation}
g^{(2)}(\lambda,\lambda^\prime)=\frac{\langle{N(\lambda)N(\lambda^{\prime})}\rangle}{\langle{N(\lambda)}\rangle\langle{N(\lambda^\prime)}\rangle}
\end{equation} 
This definition provides a direct measure of inter-wavelength intensity correlations: the diagonal  ($\lambda=\lambda^\prime$) reflects  the marginal bunching within a given spectral bin, whereas off-diagonal features reveal correlations between distinct spectral components. The measured correlation map of the driving BSV is shown in Fig. 4d. Besides the pronounced autocorrelation ridge along the diagonal, it exhibits clear exhibits symmetric off-diagonal structures, indicating the presence of correlations between different spectral modes. 
To facilitate a direct comparison of correlation strength across different wavelength pairs, we additionally define a dimensionless normalized correlation parameter 
\begin{equation}
\Gamma(\lambda,\lambda^\prime)=\frac{[g^{(2)}{(\lambda,\lambda^\prime)]^2}}{g^{(2)}(\lambda)g^{(2)}(\lambda^\prime)},
\end{equation}
where $g^{(2)}(\lambda)\equiv g^{(2)}(\lambda,\lambda)$. By construction, $\Gamma (\lambda,\lambda^\prime)$ quantifies how strongly two spectral bins co-fluctuate relative to their individual shot-to-shot fluctuations. In particular, $\Gamma$ values approaching unity indicate that the intensity fluctuations in $N(\lambda)$ and $N(\lambda^\prime)$ are highly synchronized, consistent to the reduced-noise correlations expected for a two-mode squeezed field (e.g., a reduced variance of $N(\lambda)-N(\lambda^\prime)$). Figure 4e shows the measured $\Gamma (\lambda,\lambda^\prime)$ for the multi-mode BSV: besides the diagonal elements, two symmetric off-diagonal regions also exhibit $\Gamma$ values close to unity, highlighting pairs of strongly correlated spectral components in the driving field. These structures originate from the photon-pair generation process in SPDC, i.e., the underlying BEP that imprint correlated fluctuations across the frequency components of the BSV.

\begin{figure}[p]
\centering
\includegraphics[width=1.1\textwidth]{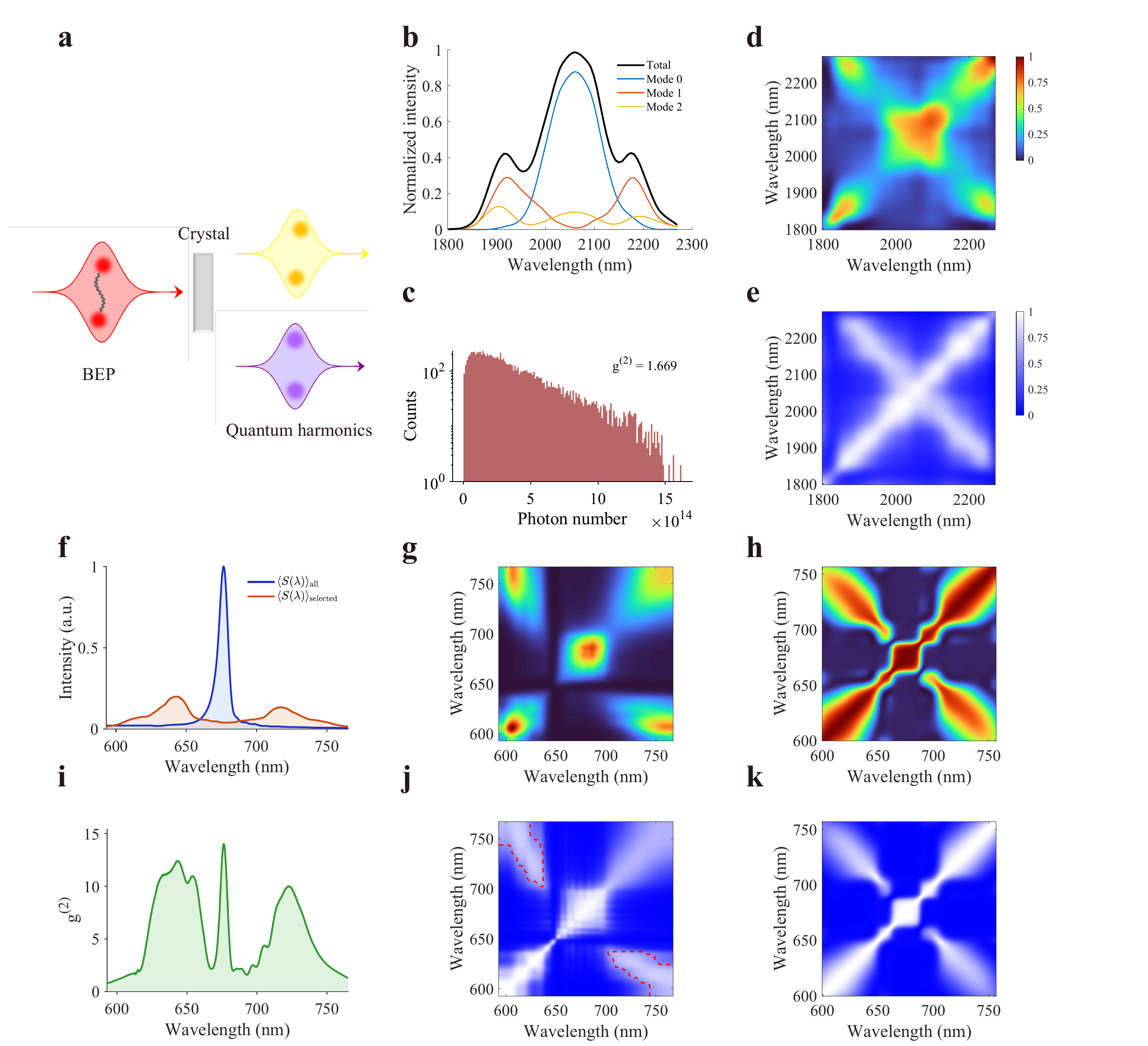}
\caption*{\small \textbf{Figure 4. Quantum-induced correlation in harmonics with BSV} \textbf{a,} Schematic illustration of HHG driven by BEP, where the quantum-induced  correlations of the input field are transferred to the emitted harmonics. \textbf{b,} Measured total spectrum of the SPDC-generated squeezed vacuum (black line), along with individual contributions from Schmidt modes (colored lines) obtained via mode decomposition. The spectral structure is mainly dominated by frequency-degenerate component (Mode 0) and non-degenerate component (Mode 1,2) . 
\textbf{c,} Photon number distribution of the multi-mode squeezed state. The reduced value $g^{(2)}=1.669$ compared to the ideal single-mode case arises from convolution over multiple Schmit modes. \textbf{d,e,} Second-order spectral correlation function $g^{(2)}(\lambda,\lambda^\prime)$ and the corresponding normalized correlation matrix $\Gamma(\lambda,\lambda^\prime)$ of the driving BSV field, showing strong off-diagonal correlations arising from spectral entanglement. \textbf{f,} Third-order harmonic total spectrum (blue) and the average of selected double-peaked pulses (orange), revealing the contributions of BEP. \textbf{i,} Wavelength-resolved  second-order correlation funcion $g^{(2)}(\lambda)$ of the third-order harmonic, showing a triple-peak pattern. This structure reflects the nonclassical imprint of both BSV and BEP features in  the harmonic emission. \textbf{g, j,} Experimentally measured $g^{(2)}(\lambda,\lambda^\prime)$ and normalized correlation matrix $\Gamma(\lambda,\lambda^\prime)$ of the third-order harmonic. The emergence of off-diagonal elements indicate that correlation of driving field  is transferred to the harmonics. \textbf{h, k,} Corresponding theoretical simulations of $g^{(2)}(\lambda,\lambda^\prime)$ and $\Gamma(\lambda,\lambda^\prime)$, which reproduce the main correlation features observed in the experiment. }
\label{fig4}
\end{figure}

\indent The natural and intriguing question that follows is whether the harmonics generated by such BEP inherit these correlations. Here, we take third-order harmonic as an example. Since Mode 0 and Mode 1 BSV dominate the driving field, the third-order harmonics are mainly generated by these two modes. Single-shot measurement allows us to distinguish the contributions from different modes by examining whether each individual harmonic pulse exhibits a single-peak or double-peak spectral structure. The overall measured spectrum of the third-order harmonic (blue curve in Fig. 4f) appears nearly single peak since it is dominated by the contribution from the single-mode (Mode 0) contribution. Meanwhile, a subset of pulses  exhibit a feature of double-peak structure that are dominated by the Mode 1 BSV, as illustrated by the red curve in Fig. 4f (individual normalized). Because higher-order Schmidt modes are intrinsically weaker than the first two degenerate modes, the gating effect of highly nonlinear HHG process further enhances the dominance of the BSV-driven component, producing the single-peaked total averaged spectrum. Nevertheless, we have examined the spectral-resolved second-order correlation $g^{(2)}$ (Fig. 4i) of third-order harmonic, which is independent of absolute intensity. A distinct triple-peak structure emerges. This clearly indicate that both single-mode and two-mode (BEP) BSV contribute to harmonic generation.   \\
\begin{figure}[p]
\centering
\includegraphics[width=1.0\textwidth]{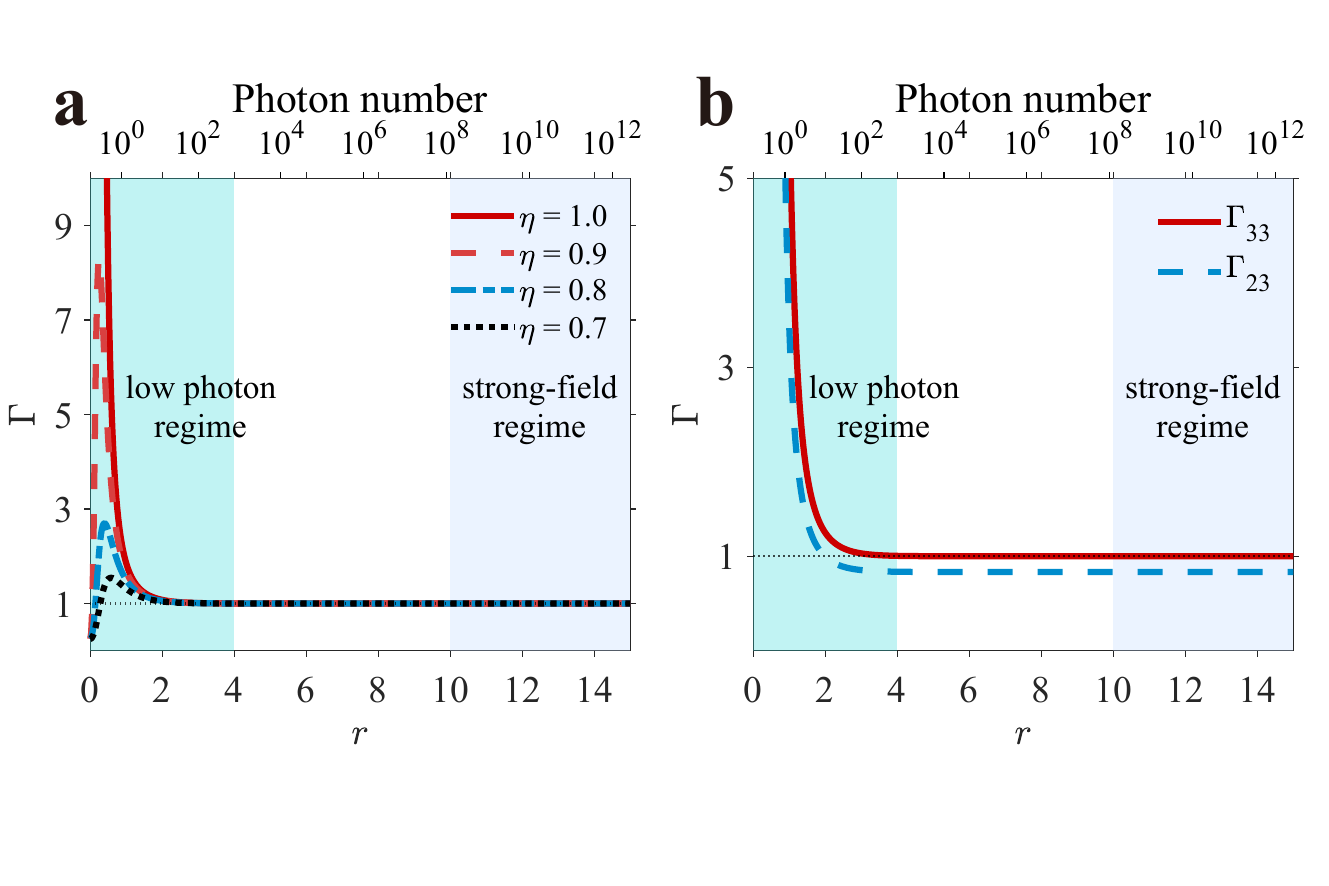}
\caption*{\small  \textbf{Figure 5. Normalized correlation parameter $\Gamma$ for a two-mode squeezed state and its nonlinear mapping in HHG} \textbf{a,} $\Gamma (r)$ of the two-mode driving field as a function of the squeezing parameter $r$, shown for different detection efficiency $\eta$. In the low photon regime, $\Gamma>1$ indicates nonclassical (pair-induced) correlations, while in the strong-field regime $\Gamma (r)$ approaches unity as the mean photon number is macroscopic, consistent with the experimentally relevant limit where the correlations manifest as nearly synchronized intensity co-fluctuations. \textbf{b,} Harmonic normalized correlation parameter $\Gamma_{mn} (r)$ versus $r$. The red and blue curves correspond to $\Gamma_{33}$ (intra-order) and $\Gamma_{23}$ (inter-order), respectively.In the macroscopic limit, $\Gamma_{33}\rightarrow1$ where as $\Gamma_{23}<1$, indicating that intra-order correlations persist while cross-order correlations are suppressed at high flux.} 
\label{fig5}
\end{figure}

To address whether the BEP-driven harmonics retain the correlations, we map the wavelength-resolved second-order cross-correlation $g^{(2)}(\lambda,\lambda^\prime)$ (Fig. 4g) together with the normalized correlation parameter $\Gamma(\lambda,\lambda^\prime)$ (Fig. 4j). Experimentally, both correlation maps exhibit-beyond the diagonal ridge-distinct symmetric off-diagonal lobes aligned with the BEP sidebands, forming a pattern strikingly similar to that observed in the driving field. This similarity indicates quantum-induced correlations between spectrally distinct components are preserved within the harmonic emission. Theoretical calculation of quantum correlation for BEP-driven HHG by solving SBE (Fig. 4h,k) reproduce these features. We also mention that under current experimental conditions, BEP contributions are limited to the second- and third-order harmonics, with no detectable signal for higher order harmonic due to the low intensity of Mode 1 (Extend Data Fig. 2). These observations show that the quantum-induced correlations can survive the highly nonlinear up-conversion process and remain  spectrally resolvable in the harmonics, enabling correlation-resolved spectroscopy of high-frequency radiation.  \\
\indent 

Our experimental observations provide unambiguous evidence of quantum-induced correlation of HHG in solids driven by quantum light. Nevertheless, it remains a question of fundamental interest to understand how the pairwise correlation of the underlying BEP are mapped through a highly nonlinear upconversion process. To capture the essential physics, we model the driving field as a two-mode squeezed state composed of modes $a$ and $b$. In the interaction representation, the Hamiltonian describing SPDC takes the form $H_{1}=\hslash g_1(ab+a^\dagger b^\dagger)$, where $g_1$ is the coupling strength and $a$, $b$ denote the non-degenerate modes at $\omega_a$, $\omega_b$, respectively. The normalized correlation parameter is then given by $\Gamma=\frac{1}{4}[1+(\frac{c}{k+\frac{1}{2}})^2]^2$, where $\eta$ is the detection efficiency of signal, $c=\eta\frac{\sinh 2r}{2}$ and $k=\eta(\frac{\cosh 2r}{2}-1)$ with $r$ being the squeezing parameter (Methods). 

Fig. 5a shows $\Gamma(r)$ for several $\eta$. In the low photon regime (small $r$), $\Gamma>1$ corresponds to a violation of the classical Cauchy-Schwarz bound, which signals a genuinely quantum origin of the correlations \cite{reid1986violations}. As $r$ increases, however, the system enters the strong-field regime where the mean photon number is macroscopic and $\Gamma (r)$ approaches unity, consistent with the experimentally relevant limit in our HHG measurements. Importantly, $\Gamma\rightarrow1$ signifies that the dominant signature becomes an almost perfectly synchronized co-fluctuation of the two spectral components.  We then analyze how the nonlinear up-conversion transfers these  correlations to the harmonic modes. In the interaction representation, we model the HHG process as $H_{2}=\hslash g_m(a^{\dagger m}c+a^mc^\dagger)+\hslash g_n(b^{\dagger n}d+b^n d^\dagger)$, where input modes at $\omega_a$ and $\omega_b$ drive the $m$th- and $n$th- harmonic modes $c$ and $d$, respectively. With this model, one can obtain the analytic form of normalized  harmonic correlation parameter,
\begin{equation}
\Gamma_{mn}=\frac{(n)!^2(m!)^2}{(2n)!(2m)!}[\sum_{k=0}^{n} \binom{n}{k}\binom{m}{n-k}\coth^{2n-2k} r]^2,
\end{equation}
We illustrate the intra-order ($\Gamma_{33}$) and inter-order ($\Gamma_{23}$) cases in Fig. 5b. In the low photon regime, $\Gamma_{mn}$ can exceed unity, indicating that the harmonic correlations retain a distinctly nonclassical feature. In the strong-field regime, however, the behavior depends crucially on whether the two harmonics originate from the matched conversion process. For the intra-order case $m=n$, $\Gamma _{mm}(r)$ approaches unity, indicating that the pair induced correlation survives as an almost perfectly synchronized co-fluctuation. By contrast, for $m\neq n$ the inter-order correlation weakens in the macroscopic limit and $\Gamma_{mn} (r)$ drops below unity. Taken together, these results show that under strong-field conditions, the experimentally relevant signature is a near-unity $\Gamma$ for $m=n$ accompanied by reduced noise in correlated photon number, while cross-order correlations ($m\neq n$) are progressively washed out.  \\
\indent In conclusion, we have demonstrated that HHG in solids driven by single-mode and two-mode BSV provides a correlation-preserving route for accessing the strong-field regime  with quantum light. Utilizing bright squeezed vacuum, we generated high-order harmonics up to the 10th order and established two central results. 

First, under single-mode BSV driving, the harmonic emission exhibits strong superbunching with a non-monotonic $g^{(2)}$ versus order: it rises from low orders, peaks at intermediate orders, and decreases at high orders. This behavior is captured by quantum-corrected semiconductor Bloch simulations. Importantly, the observed photon statistics not only signal nonclassicality but also serve as a diagnostic of intraband and interband emission channels, highlighting quantum-optical analysis of HHG as a sensitive probe of microscopic electron dynamics. 

Second, when non-degenerate entangled photon pairs are present in the driving field, the harmonics retain quantum-induced correlations that are spectrally resolvable. Both of wavelength-resolved cross-correlation map $g^{(2)}(\lambda,\lambda^\prime)$ and the normalized correlation parameter $\Gamma(\lambda,\lambda^\prime)$ reveal symmetric off-diagonal correlation lobes at the harmonic wavelengths corresponding to the BEP. Notably, these features persist even in the macroscopic photon number regime, where $\Gamma\to 1$ due to intensity co-fluctuation. A fully quantum model further yields closed-form expressions for $\Gamma_{mn}$ between harmonic orders. This model shows that intra-order correlations ($m=n$) survive in the macroscopic limit, while cross-order correlations ($m\neq n$) are progressively suppressed at high flux. \\
\indent Beyond validating the quantum character of solid-state HHG, our results establish correlation-resolved HHG as a methodology: intensity-normalized metrics such as $g^{(2)}$ and $\Gamma$ constitute a form of  statistical spectroscopy that complements the intensity spectra and provides access to information encoded in fluctuations and correlations. This modality establishes HHG as a practical interface for mapping the correlations into high-frequency radiation, opening avenues for quantum-enhanced attosecond spectroscopy, correlation-based metrology, and nonclassical XUV sources. Looking forward, phase-sensitive detection in harmonic measurements-via harmonic homodyne/heterodyne in the XUV will move beyond second-order metrics toward quantum tomography of the harmonic field. Extending these tools to strongly correlated and topological solids could enable entanglement-aware probes of many-body electron dynamics and pave the way for the control of quantum light-matter interactions and the development of compact, correlation-engineered attosecond sources.

\section*{Methods}
\section*{Experimental methods}
\noindent\textbf{Single-mode and two-mode bright squeezed light generation}.\\
In the experiment, near-infrared laser pulses (1030 nm, 2 mJ) were delivered from a commercial femtosecond Yb:YAG fiber laser with  $140$ fs duration and a tunable repetition rate of $1$ Hz - $10$ kHz. A portion of the laser beam was reduced in beam diameter and sent into two 3 mm thick BBO crystals placed in tandem to generate multimode BSV centered at 2060 nm via  type-I SPDC. A crystal spacing of approximately $40$ cm was introduced to select the first spatial mode by ensuring phase matching for the lowest-divergence component. Without spectral filtering, the resulting BSV contains a few lower-order spectral Schmidt modes. Without damaging the crystals, the average single pulse energy of the generated multi-mode BSV reached $2 \ \mathrm{\mu J}$. To isolate a single spectral mode, a band-pass filter centered at $2050\ \mathrm{nm}$ with  a FWHM  of $100\ \mathrm{nm}$ was applied, yielding a single-mode BSV with $g^{(2)}=2.655$. Its average pulse energy is  ~$\sim 0.8 \mathrm{\mu J}$, corresponding to approximately $8.3\times 10^{12}$ photons and a squeezing parameter of $r=15.6$. The single pulse spectrum of the driving field was characterized using a near-infrared spectrometer (AvaSpec-NIR256-2.5-HSC-EVO). The quantum light (either multi-mode or filtered single-mode BSV) was then focused by an off-axis parabolic mirror (OAP) ($f=25.4 \ \mathrm{mm}$) into a $200 \ \mathrm{\mu m}$-thick free-standing $x$-cut lithium niobate (LN) crystal, with the laser polarization aligned along the crystal's $\Gamma-\mathrm{Z}$ axis. The generated harmonics were collected by a matching OAP and measured using a Princeton Instruments HRS-500 spectrometer equipped with a high efficiency detector (PIX-100B) for single-pulse-resolved spectra.

\noindent\textbf{Schmidt mode decomposition of quantum light}.\\ 
 To characterize the mode structure of BSV, we analyze over 10000 single-shot spectra $I(\lambda)$. These spectra encode both amplitude fluctuations and spectral correlations of the quantum field. The intensity covariance matrix across multiple pulses is defined as  \cite{barakat2025simultaneous}:
\begin{equation}
\mathrm{Cov} (\lambda,\lambda^\prime)=\langle I(\lambda) I(\lambda^\prime)\rangle-\langle I(\lambda)\rangle\langle I(\lambda^\prime)\rangle
\end{equation}
We then normalize it to obtain the dimensionless coherence function,
 \begin{equation}
 C(\lambda,\lambda^\prime)= \frac{\mathrm{Cov}(\lambda,\lambda^\prime)}{\iint \mathrm{Cov}(\lambda,\lambda^\prime)d\lambda d\lambda^\prime}
 \end{equation}
We then apply singular value decomposition (SVD) to the square root of the matrix to extract a set of orthonormal spectral modes $u_m(\lambda)$ and their corresponding weights $\lambda_m$, such that:
\begin{equation}
C(\lambda,\lambda^\prime)=|\sum_m\lambda_mu_m(\lambda)u^*_m(\lambda^\prime)|^2
\end{equation}
The weights $\lambda_m$ represent the normalized mode populations. To visualize each mode's contribution to the total spectrum, we plot the intensity distribution $\lambda_m |u_m(\lambda)|^2$, which combines both the mode shape and its relative weight (Fig. 2a, 4b).

\section*{Theoretical methods}
\noindent\textbf{Semiconductor Bloch Equation Simulation  }. 
We simulate HHG in solids using the semiconductor Bloch equations (SBE) within a two-band model . The evolution of the interband coherence $\pi (K,t)$ and the band populations $n_m(K,t)$ (with $m=v,c$ for valence and conduction bands, respectively) is governed by:
\begin{equation}
\begin{aligned}
&&\dot{\pi}(\mathbf{K},t)=-\frac{\pi(\mathbf{K},t)}{T_2}-i\Omega (\mathbf{K},t)w(\mathbf{K},t)e^{-iS(\mathbf{K},t)} \\
&&\dot{n}(\mathbf{K},t)=is_m\Omega^*(\mathbf{K},t) \pi (\mathbf{K},t)e^{iS(\mathbf{K},t)}+c.c.
\end{aligned}
\end{equation}
Here, $w=n_v-n_c$ is the population inversion, and $T_2$ is the dephasing time (set to $T_0/5$ in our simulations), and $s_m=+1$ for $m=c$, $-1$ for $m=v$. The momentum $\mathbf{K}=\mathbf{k}-\mathbf{A}(t)$ is defined in the moving frame under the vector potential $\mathbf{A}(t)$. The Rabi frequency is given by  ${\Omega}(\mathbf{K},t)=\mathbf{F}(t)\cdot\mathbf{d}[\mathbf{K}+\mathbf{A}(t)]$, where
$\mathbf{d(\mathbf{k})}$ is the interband dipole matrix element and $\mathbf{F}(t)$ is the electric field. The classical action is defined as $S=\int_{-\infty}^t E_g[\mathbf{K}+\mathbf{A}(t^\prime)]dt^\prime$. The effective bandgap $E_g(\mathbf{F})$ incorporates the contribution from spontaneous polarization via a Stark-like shift, and is expressed as $E_g(\mathbf{F})=E_g(0)-\mathbf{\mu}\cdot\mathbf{F}(t)$, where $E_g(0)$ is the field-free bandgap energy, taken to be $3.3$ eV \cite{friedrich2017polaron,shao2022spontaneous}. Here $\mathbf{\mu}$ is the electric dipole determined from the spontaneous polarization of LN. The interband polarization is written as: $p(\mathbf{K},t)=\mathbf{d}(\mathbf{K+\mathbf{A}})\pi (\mathbf{K},t)e^{iS(\mathbf{K},t)}+c.c.$. The total current consists of interband and intraband contributions: $\mathbf{j}_{\mathrm{ra}}(t)=\sum_{m=c,v}\int_{\mathrm{BZ}}\mathbf{v}_m[\mathbf{K}+\mathbf{A}(t)]n_m(\mathbf{K},t)d^3 \mathbf{K}$ and $\mathbf{j}_{\mathrm{er}}(t)=\frac{d}{dt}\int_{\mathrm{BZ}}\mathrm{p}(\mathrm{K},t)d^3\mathrm{K}$.  The HHG spectrum  under a coherent field is calculated as : $S_c(\omega)=|\int_{-\infty}^{\infty}[\mathbf{j}_{\mathrm{er}}(t)+\mathbf{j}_{\mathrm{ra}}(t)]e^{i\omega t}dt|^2$. To incorporate the quantum nature of the BSV, we adopt the Husimi Q-function formalism. Each BSV pulse is treated as a statistical mixture of coherent states with field amplitude $\varepsilon_{\alpha}$, distributed according to $Q(\varepsilon_\alpha)=\frac{2}{\pi |\varepsilon_0|^2}\exp (-\frac{|\varepsilon_\alpha|^2}{2|\varepsilon_0|^2})$, where $\varepsilon_0$ is the average amplitude of the BSV field. The overall HHG spectrum is obtained by averaging the coherent spectra weight by the Q-function: $S_q(\omega)=\int Q(\varepsilon_\alpha)S_c(\omega,\varepsilon_\alpha)d\varepsilon_\alpha$. To extract photon number statistics for each harmonic order, we  analyze the distribution of harmonic photon numbers $N_{n\omega}$ across the set of simulated BSV-driven pulses. The probability of  $N_{n\omega}$ photons is given by $P(N_{n\omega})=P_{\alpha}(N_{\alpha})(\frac{d N_{\alpha}}{d N_{n\omega}})$, where $P_{\alpha}(N_{\alpha})$ is the photon number distribution of the BSV, and the Jacobian $d N_{\alpha}/d N_{n\omega}$ accounts for the nonlinear mapping between input and output photon numbers. This formulation enables the direct evaluation of high-order correlation functions, such as $g^{(2)}$, from the simulated data.  

\noindent\textbf{Entanglement in harmonic emission}. In our experiment, the driving field is a two-mode squeezed state that is generated from SPDC with $H_{1}=\hslash g_1(ab+a^\dagger b^\dagger)$. Since it is a Gaussian state, it can be fully described by its covariance matrix\cite{duan2000inseparability,serafini2023quantum}, 
\begin{equation}
\sigma_{A B}=\left(\begin{array}{cccc}
n & 0 & c_1 & 0 \\
0 & n & 0 & c_2 \\
c_1 & 0 & m & 0 \\
0 & c_2 & 0 & m
\end{array}\right).
\end{equation}
Here, $n=\mathrm{Var}(x_A)=\mathrm{Var}(p_A)$, $m=\mathrm{Var}(x_B)=\mathrm{Var}(p_B)$ represent the variance of the amplitude and phase quadratures, $x_A=(a+a^\dagger)/\sqrt{2}$, $x_B=(b+b^\dagger)/\sqrt{2}$, $p_A=(a-a^\dagger)/\sqrt{2}i$, and $p_B=(b-b^\dagger)/\sqrt{2}i$. $c_1=\mathrm{Cov}(x_A,x_B)$ and $c_2=\mathrm{Cov}(p_A,p_B)$ indicate their cross correlations. In ideal cases, the correlated variances between two-mode squeezed state can be obtained directly as $\mathrm{Var}(x_A+x_B)=\mathrm{Var}(p_A-p_B)=e^{2r}\equiv 2V_{+}$ and $\mathrm{Var}(x_A-x_B)=\mathrm{Var}(p_A+p_B)=e^{-2r}\equiv2V_{-}$, where $r=g_1t$ is the effective squeezing parameter. Taking into account the practical losses that can be modeled by mixing the mode with an auxiliary vacuum on a beam splitter with transmissivity $\eta$\cite{giovannetti2014ultimate,mari2014quantum}, the elements of the covariance matrix become $n=m=\eta(V_{+}+V_{-})/2+1-\eta$, $c_1=-c_2=\eta(V_{+}-V_{-})/2$. Therefore, the normalized correlation parameter is derived as $\Gamma=\frac{1}{4}[1+\frac{c_1^2}{(n-\frac{1}{2})(m-\frac{1}{2})}]^2$ by using the relation between moments in the Gaussian distribution. Defining $c=\eta\frac{\sinh 2r}{2}$ and $k=\eta(\frac{\cosh 2r}{2}-1)$, normalized correlation parameter is then given by $\Gamma=\frac{1}{4}[1+(\frac{c}{k+\frac{1}{2}})^2]^2$. In addition, for harmonic modes we make use of the up-conversion Hamiltonian in the interaction picture, which takes the form of $H_{2}=\hslash g_m(a^{\dagger m}c+a^mc^\dagger)+\hslash g_n(b^{\dagger n}d+b^n d^\dagger)$. It is noticed that in the interaction picture, the operators satisfy $c_I(t)=ce^{-i\omega_ct}$ and $d_I(t)=de^{-i\omega_dt}$, while the state vector follows $|\psi(t)\rangle=e^{-iH_It/\hslash}|\psi(0)\rangle$. Thus, directly inserting the Hamiltonian with weak nonlinear coupling strengths $g_{n(m)}\ll1$, we will get $\langle c^\dagger c\rangle=m!g_m^2\sinh^{2m}(r)$,  $\langle c^{\dagger2} c^2\rangle=(2m)!g_m^4\sinh^{4m}(r)$, $\langle d^\dagger d\rangle=n!g_n^2\sinh^{2n}(r)$,  $\langle d^{\dagger2} d^2\rangle=(2n)!g_n^4\sinh^{4n}(r)$, and $\langle c^\dagger d^\dagger cd \rangle=m!n!g_m^2g_n^2\sinh^{2m+2n}r\sum_{k=0}^n\binom{n}{k}\binom{m}{n-k}\coth^{2n-2k}r$. Here, we have assumed $n\leq m$. And the normalized correlation parameter between the $m$th and $n$th harmonic modes is then obtained as $\Gamma_{mn}=\frac{[g_{mn}^{(2)}]^2}{g^{(2)}_m g^{(2)}_n}=\frac{(n)!^2(m!)^2}{(2n)!(2m)!}[\sum_{k=0}^{n} \binom{n}{k}\binom{m}{n-k}\coth^{2n-2k} r]^2$.

\bibliography{sn-bibliography}

\begin{thebibliography}{10}
\expandafter\ifx\csname url\endcsname\relax
  \def\url#1{\burl{#1}}\fi
\expandafter\ifx\csname urlprefix\endcsname\relax\def\urlprefix{URL }\fi
\providecommand{\bibinfo}[2]{#2}
\providecommand{\eprint}[2][]{\url{#2}}
\providecommand{\doi}[1]{\url{https://doi.org/#1}}
\bibcommenthead

\bibitem{glauber1963coherent}
\bibinfo{author}{Glauber, R.~J.}
\newblock \bibinfo{title}{Coherent and incoherent states of the radiation field}.
\newblock \emph{\bibinfo{journal}{Physical Review}} \textbf{\bibinfo{volume}{131}}, \bibinfo{pages}{2766} (\bibinfo{year}{1963}).

\bibitem{scully1997quantum}
\bibinfo{author}{Scully, M.~O.} \& \bibinfo{author}{Zubairy, M.~S.}
\newblock \emph{\bibinfo{title}{Quantum optics}}  (\bibinfo{publisher}{Cambridge university press}, \bibinfo{year}{1997}).

\bibitem{grynberg2010introduction}
\bibinfo{author}{Grynberg, G.}, \bibinfo{author}{Aspect, A.} \& \bibinfo{author}{Fabre, C.}
\newblock \emph{\bibinfo{title}{Introduction to quantum optics: from the semi-classical approach to quantized light}}  (\bibinfo{publisher}{Cambridge university press}, \bibinfo{year}{2010}).

\bibitem{loudon2000quantum}
\bibinfo{author}{Loudon, R.}
\newblock \emph{\bibinfo{title}{The quantum theory of light}}  (\bibinfo{publisher}{OUP Oxford}, \bibinfo{year}{2000}).

\bibitem{ferray1988multiple}
\bibinfo{author}{Ferray, M.} \emph{et~al.}
\newblock \bibinfo{title}{Multiple-harmonic conversion of 1064 nm radiation in rare gases}.
\newblock \emph{\bibinfo{journal}{Journal of Physics B: Atomic, Molecular and Optical Physics}} \textbf{\bibinfo{volume}{21}}, \bibinfo{pages}{L31} (\bibinfo{year}{1988}).

\bibitem{mcpherson1987studies}
\bibinfo{author}{McPherson, A.}, \bibinfo{author}{Gibson, G.}, \bibinfo{author}{Jara, H.}, \bibinfo{author}{Johann, U.} \& \bibinfo{author}{McIntyre, I.}
\newblock \bibinfo{title}{Studies of multiphoton production of vacuum-ultraviolet radiation in the rare gases}.
\newblock \emph{\bibinfo{journal}{Journal of the Optical Society of America B}} \textbf{\bibinfo{volume}{4}}, \bibinfo{pages}{595--601} (\bibinfo{year}{1987}).

\bibitem{lewenstein1994theory}
\bibinfo{author}{Lewenstein, M.}, \bibinfo{author}{Balcou, P.}, \bibinfo{author}{Ivanov, M.~Y.}, \bibinfo{author}{L’huillier, A.} \& \bibinfo{author}{Corkum, P.~B.}
\newblock \bibinfo{title}{Theory of high-harmonic generation by low-frequency laser fields}.
\newblock \emph{\bibinfo{journal}{Physical Review A}} \textbf{\bibinfo{volume}{49}}, \bibinfo{pages}{2117} (\bibinfo{year}{1994}).

\bibitem{corkum1993plasma}
\bibinfo{author}{Corkum, P.~B.}
\newblock \bibinfo{title}{Plasma perspective on strong field multiphoton ionization}.
\newblock \emph{\bibinfo{journal}{Physical Review Letters}} \textbf{\bibinfo{volume}{71}}, \bibinfo{pages}{1994} (\bibinfo{year}{1993}).

\bibitem{schafer1993above}
\bibinfo{author}{Schafer, K.}, \bibinfo{author}{Yang, B.}, \bibinfo{author}{DiMauro, L.} \& \bibinfo{author}{Kulander, K.}
\newblock \bibinfo{title}{Above threshold ionization beyond the high harmonic cutoff}.
\newblock \emph{\bibinfo{journal}{Physical review letters}} \textbf{\bibinfo{volume}{70}}, \bibinfo{pages}{1599} (\bibinfo{year}{1993}).

\bibitem{tsatrafyllis2017high}
\bibinfo{author}{Tsatrafyllis, N.}, \bibinfo{author}{Kominis, I.}, \bibinfo{author}{Gonoskov, I.} \& \bibinfo{author}{Tzallas, P.}
\newblock \bibinfo{title}{High-order harmonics measured by the photon statistics of the infrared driving-field exiting the atomic medium}.
\newblock \emph{\bibinfo{journal}{Nature Communications}} \textbf{\bibinfo{volume}{8}}, \bibinfo{pages}{15170} (\bibinfo{year}{2017}).

\bibitem{tsatrafyllis2019quantum}
\bibinfo{author}{Tsatrafyllis, N.} \emph{et~al.}
\newblock \bibinfo{title}{Quantum optical signatures in a strong laser pulse after interaction with semiconductors}.
\newblock \emph{\bibinfo{journal}{Physical Review Letters}} \textbf{\bibinfo{volume}{122}}, \bibinfo{pages}{193602} (\bibinfo{year}{2019}).

\bibitem{lewenstein2021generation}
\bibinfo{author}{Lewenstein, M.} \emph{et~al.}
\newblock \bibinfo{title}{Generation of optical schr{\"o}dinger cat states in intense laser--matter interactions}.
\newblock \emph{\bibinfo{journal}{Nature Physics}} \textbf{\bibinfo{volume}{17}}, \bibinfo{pages}{1104--1108} (\bibinfo{year}{2021}).

\bibitem{theidel2024evidence}
\bibinfo{author}{Theidel, D.} \emph{et~al.}
\newblock \bibinfo{title}{Evidence of the quantum optical nature of high-harmonic generation}.
\newblock \emph{\bibinfo{journal}{PRX Quantum}} \textbf{\bibinfo{volume}{5}}, \bibinfo{pages}{040319} (\bibinfo{year}{2024}).

\bibitem{theidel2025observation}
\bibinfo{author}{Theidel, D.} \emph{et~al.}
\newblock \bibinfo{title}{Observation of a displaced squeezed state in high-harmonic generation}.
\newblock \emph{\bibinfo{journal}{Physical Review Research}} \textbf{\bibinfo{volume}{7}}, \bibinfo{pages}{033223} (\bibinfo{year}{2025}).

\bibitem{lamprou2025nonlinear}
\bibinfo{author}{Lamprou, T.}, \bibinfo{author}{Rivera-Dean, J.}, \bibinfo{author}{Stammer, P.}, \bibinfo{author}{Lewenstein, M.} \& \bibinfo{author}{Tzallas, P.}
\newblock \bibinfo{title}{Nonlinear optics using intense optical coherent state superpositions}.
\newblock \emph{\bibinfo{journal}{Physical Review Letters}} \textbf{\bibinfo{volume}{134}}, \bibinfo{pages}{013601} (\bibinfo{year}{2025}).

\bibitem{stammer2024entanglement}
\bibinfo{author}{Stammer, P.} \emph{et~al.}
\newblock \bibinfo{title}{Entanglement and squeezing of the optical field modes in high harmonic generation}.
\newblock \emph{\bibinfo{journal}{Physical Review Letters}} \textbf{\bibinfo{volume}{132}}, \bibinfo{pages}{143603} (\bibinfo{year}{2024}).

\bibitem{rivera2024squeezed}
\bibinfo{author}{Rivera-Dean, J.} \emph{et~al.}
\newblock \bibinfo{title}{Squeezed states of light after high-order harmonic generation in excited atomic systems}.
\newblock \emph{\bibinfo{journal}{Physical Review A}} \textbf{\bibinfo{volume}{110}}, \bibinfo{pages}{063118} (\bibinfo{year}{2024}).

\bibitem{yi2025generation}
\bibinfo{author}{Yi, S.} \emph{et~al.}
\newblock \bibinfo{title}{Generation of massively entangled bright states of light during harmonic generation in resonant media}.
\newblock \emph{\bibinfo{journal}{Physical Review X}} \textbf{\bibinfo{volume}{15}}, \bibinfo{pages}{011023} (\bibinfo{year}{2025}).

\bibitem{gonoskov2024nonclassical}
\bibinfo{author}{Gonoskov, I.} \emph{et~al.}
\newblock \bibinfo{title}{Nonclassical light generation and control from laser-driven semiconductor intraband excitations}.
\newblock \emph{\bibinfo{journal}{Physical Review B}} \textbf{\bibinfo{volume}{109}}, \bibinfo{pages}{125110} (\bibinfo{year}{2024}).

\bibitem{pizzi2023light}
\bibinfo{author}{Pizzi, A.}, \bibinfo{author}{Gorlach, A.}, \bibinfo{author}{Rivera, N.}, \bibinfo{author}{Nunnenkamp, A.} \& \bibinfo{author}{Kaminer, I.}
\newblock \bibinfo{title}{Light emission from strongly driven many-body systems}.
\newblock \emph{\bibinfo{journal}{Nature Physics}} \textbf{\bibinfo{volume}{19}}, \bibinfo{pages}{551--561} (\bibinfo{year}{2023}).

\bibitem{lange2024electron}
\bibinfo{author}{Lange, C.~S.}, \bibinfo{author}{Hansen, T.} \& \bibinfo{author}{Madsen, L.~B.}
\newblock \bibinfo{title}{Electron-correlation-induced nonclassicality of light from high-order harmonic generation}.
\newblock \emph{\bibinfo{journal}{Physical Review A}} \textbf{\bibinfo{volume}{109}}, \bibinfo{pages}{033110} (\bibinfo{year}{2024}).

\bibitem{lange2025excitonic}
\bibinfo{author}{Lange, C.~S.}, \bibinfo{author}{Hansen, T.} \& \bibinfo{author}{Madsen, L.~B.}
\newblock \bibinfo{title}{Excitonic enhancement of squeezed light in quantum-optical high-harmonic generation from a mott insulator}.
\newblock \emph{\bibinfo{journal}{Physical Review Letters}} \textbf{\bibinfo{volume}{135}}, \bibinfo{pages}{043603} (\bibinfo{year}{2025}).

\bibitem{even2023photon}
\bibinfo{author}{Even~Tzur, M.} \emph{et~al.}
\newblock \bibinfo{title}{Photon-statistics force in ultrafast electron dynamics}.
\newblock \emph{\bibinfo{journal}{Nature Photonics}} \textbf{\bibinfo{volume}{17}}, \bibinfo{pages}{501--509} (\bibinfo{year}{2023}).

\bibitem{gorlach2023high}
\bibinfo{author}{Gorlach, A.} \emph{et~al.}
\newblock \bibinfo{title}{High-harmonic generation driven by quantum light}.
\newblock \emph{\bibinfo{journal}{Nature Physics}} \textbf{\bibinfo{volume}{19}}, \bibinfo{pages}{1689--1696} (\bibinfo{year}{2023}).

\bibitem{sh2012superbunched}
\bibinfo{author}{Iskhakov, T.~S.}, \bibinfo{author}{P{\'e}rez, A.}, \bibinfo{author}{Spasibko, K.~Y.}, \bibinfo{author}{Chekhova, M.} \& \bibinfo{author}{Leuchs, G.}
\newblock \bibinfo{title}{Superbunched bright squeezed vacuum state}.
\newblock \emph{\bibinfo{journal}{Optics letters}} \textbf{\bibinfo{volume}{37}}, \bibinfo{pages}{1919--1921} (\bibinfo{year}{2012}).

\bibitem{iskhakov2012polarization}
\bibinfo{author}{Iskhakov, T.~S.}, \bibinfo{author}{Agafonov, I.~N.}, \bibinfo{author}{Chekhova, M.~V.} \& \bibinfo{author}{Leuchs, G.}
\newblock \bibinfo{title}{Polarization-entangled light pulses of 10$^5$ photons}.
\newblock \emph{\bibinfo{journal}{Physical Review Letters}} \textbf{\bibinfo{volume}{109}}, \bibinfo{pages}{150502} (\bibinfo{year}{2012}).

\bibitem{iskhakov2016heralded}
\bibinfo{author}{Iskhakov, T.~S.}
\newblock \bibinfo{title}{Heralded source of bright multi-mode mesoscopic sub-poissonian light}.
\newblock \emph{\bibinfo{journal}{Optics letters}} \textbf{\bibinfo{volume}{41}}, \bibinfo{pages}{2149--2152} (\bibinfo{year}{2016}).

\bibitem{rasputnyi2024high}
\bibinfo{author}{Rasputnyi, A.} \emph{et~al.}
\newblock \bibinfo{title}{High-harmonic generation by a bright squeezed vacuum}.
\newblock \emph{\bibinfo{journal}{Nature Physics}} \textbf{\bibinfo{volume}{20}}, \bibinfo{pages}{1960--1965} (\bibinfo{year}{2024}).

\bibitem{lemieux2025photon}
\bibinfo{author}{Lemieux, S.} \emph{et~al.}
\newblock \bibinfo{title}{Photon bunching in high-harmonic emission controlled by quantum light}.
\newblock \emph{\bibinfo{journal}{Nature Photonics}} \bibinfo{pages}{1--5} (\bibinfo{year}{2025}).

\bibitem{tzur2025measuring}
\bibinfo{author}{Tzur, M.~E.} \emph{et~al.}
\newblock \bibinfo{title}{Measuring and controlling the birth of quantum attosecond pulses}.
\newblock \emph{\bibinfo{journal}{arXiv preprint arXiv:2502.09427}}  (\bibinfo{year}{2025}).

\bibitem{spasibko2017multiphoton}
\bibinfo{author}{Spasibko, K.~Y.} \emph{et~al.}
\newblock \bibinfo{title}{Multiphoton effects enhanced due to ultrafast photon-number fluctuations}.
\newblock \emph{\bibinfo{journal}{Physical Review Letters}} \textbf{\bibinfo{volume}{119}}, \bibinfo{pages}{223603} (\bibinfo{year}{2017}).

\bibitem{glauber1963quantum}
\bibinfo{author}{Glauber, R.~J.}
\newblock \bibinfo{title}{The quantum theory of optical coherence}.
\newblock \emph{\bibinfo{journal}{Physical Review}} \textbf{\bibinfo{volume}{130}}, \bibinfo{pages}{2529} (\bibinfo{year}{1963}).

\bibitem{reid1986violations}
\bibinfo{author}{Reid, M.} \& \bibinfo{author}{Walls, D.}
\newblock \bibinfo{title}{Violations of classical inequalities in quantum optics}.
\newblock \emph{\bibinfo{journal}{Physical Review A}} \textbf{\bibinfo{volume}{34}}, \bibinfo{pages}{1260} (\bibinfo{year}{1986}).

\bibitem{barakat2025simultaneous}
\bibinfo{author}{Barakat, I.} \emph{et~al.}
\newblock \bibinfo{title}{Simultaneous measurement of multimode squeezing through multimode phase-sensitive amplification}.
\newblock \emph{\bibinfo{journal}{Optica Quantum}} \textbf{\bibinfo{volume}{3}}, \bibinfo{pages}{36--44} (\bibinfo{year}{2025}).

\bibitem{friedrich2017polaron}
\bibinfo{author}{Friedrich, M.}, \bibinfo{author}{Schmidt, W.~G.}, \bibinfo{author}{Schindlmayr, A.} \& \bibinfo{author}{Sanna, S.}
\newblock \bibinfo{title}{Polaron optical absorption in congruent lithium niobate from time-dependent density-functional theory}.
\newblock \emph{\bibinfo{journal}{Physical Review Materials}} \textbf{\bibinfo{volume}{1}} (\bibinfo{year}{2017}).

\bibitem{shao2022spontaneous}
\bibinfo{author}{Shao, T.-J.}, \bibinfo{author}{Hu, F.} \& \bibinfo{author}{Chen, H.-B.}
\newblock \bibinfo{title}{Spontaneous polarization effects on solid high harmonic generation in ferroelectric lithium niobate crystals}.
\newblock \emph{\bibinfo{journal}{Journal of Physics B: Atomic, Molecular and Optical Physics}} \textbf{\bibinfo{volume}{54}}, \bibinfo{pages}{245402} (\bibinfo{year}{2022}).

\bibitem{duan2000inseparability}
\bibinfo{author}{Duan, L.-M.}, \bibinfo{author}{Giedke, G.}, \bibinfo{author}{Cirac, J.~I.} \& \bibinfo{author}{Zoller, P.}
\newblock \bibinfo{title}{Inseparability criterion for continuous variable systems}.
\newblock \emph{\bibinfo{journal}{Physical Review Letters}} \textbf{\bibinfo{volume}{84}}, \bibinfo{pages}{2722} (\bibinfo{year}{2000}).

\bibitem{serafini2023quantum}
\bibinfo{author}{Serafini, A.}
\newblock \emph{\bibinfo{title}{Quantum continuous variables: a primer of theoretical methods}}  (\bibinfo{publisher}{CRC press}, \bibinfo{year}{2023}).

\bibitem{giovannetti2014ultimate}
\bibinfo{author}{Giovannetti, V.}, \bibinfo{author}{Garcia-Patron, R.}, \bibinfo{author}{Cerf, N.~J.} \& \bibinfo{author}{Holevo, A.~S.}
\newblock \bibinfo{title}{Ultimate classical communication rates of quantum optical channels}.
\newblock \emph{\bibinfo{journal}{Nature Photonics}} \textbf{\bibinfo{volume}{8}}, \bibinfo{pages}{796--800} (\bibinfo{year}{2014}).

\bibitem{mari2014quantum}
\bibinfo{author}{Mari, A.}, \bibinfo{author}{Giovannetti, V.} \& \bibinfo{author}{Holevo, A.~S.}
\newblock \bibinfo{title}{Quantum state majorization at the output of bosonic gaussian channels}.
\newblock \emph{\bibinfo{journal}{Nature Communications}} \textbf{\bibinfo{volume}{5}}, \bibinfo{pages}{3826} (\bibinfo{year}{2014}).

\end{thebibliography}

\section*{Author Contributions} 
 Z. L. designed the experiments and conducted SBE simulations. F. S. built the theoretical model of harmonic generation by two-mode squeezed vacuum state. Z. L., J. L., H. L.  performed the experiments. Z. L., F. S., S.Y., M.I. and Y. L. analyzed the data.  Z. L., M.I. and Y. L. drafted the paper with extensive input from all authors. M. I. and Y. L. conceived the idea and supervised the project. All authors provided critical feedback and helped shape the research, analysis and manuscript.

\section*{Data availability}
The data that support the plots within this paper and other findings of this study is available from the corresponding author upon reasonable request.

\section*{Code availability}
The code used to produce the results are available from the corresponding author upon reasonable request.

\section*{Acknowledgements}

This work was supported by the National Key R$\&$D Program (Nos. 2022YFA1604301 and 2023YFA1406803) and Natural Science Foundation of China (Nos. 12334013, 92050201, 92250306, and 12474256). M.I. and S.Y. acknowledge the European Union’s Horizon Europe research and innovation programme under the Marie Skłodowska-Curie grant agreement No. 101168628 (project QU-ATTO) and the DFG grant IV 152/11-1, Project number 545591821.

\section*{Additional information}
 Correspondence and requests for materials should be addressed to Y. L.

\begin{figure}[ht]
\centering
\includegraphics[width=1\textwidth]{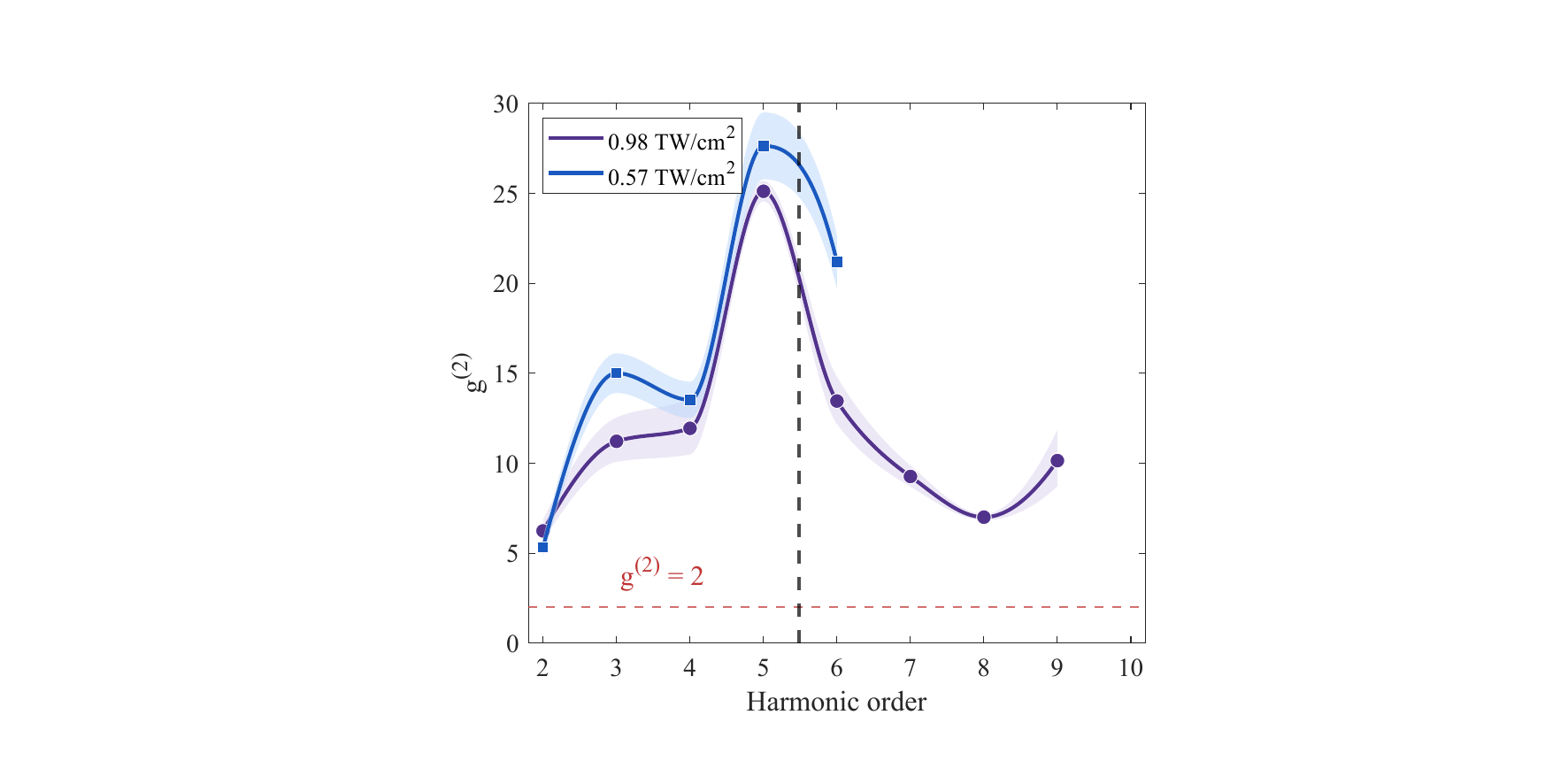}
\caption*{\small \textbf{Extended Figure 1.} \textbf{ Robust nonmonotonic trend of $g^{(2)}$ versus harmonic order. }  Second-order correlation function $g^{(2)}$ measured at two average driving intensity (0.57 and 0.98 TW/cm$^2$), demonstrating that the increase-decrease trend is preserved across different intensities.      }
\label{efig1}
\end{figure}
\begin{figure}[ht]
\centering
\includegraphics[width=1\textwidth]{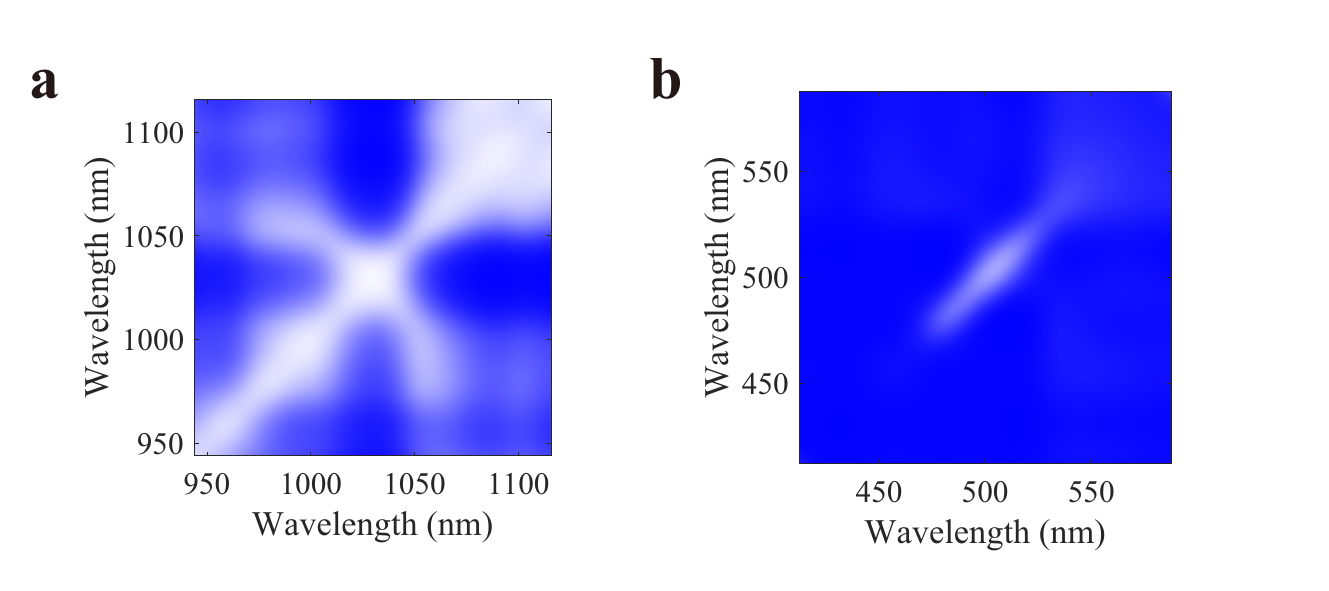}
\caption*{\small \textbf{Extended Figure 2.} Normalized correlation parameter $\Gamma(\lambda,\lambda^\prime)$ under multi-mode BSV driving of second- and forth-order harmonic. Pronounced off-diagonal correlations are observed in the second-order harmonic, while the fourth harmonic exhibits only diagonal feature. This indicates that BEP  predominantly generates second- and third-order harmonics, without contribution to the fourth harmonics.}
\label{efig2}
\end{figure}

\end{document}